\documentstyle[aps,prb,twocolumn,epsfig]{revtex}
\begin{document}
\flushbottom
\twocolumn[
\hsize\textwidth\columnwidth\hsize\csname @twocolumnfalse\endcsname

\title{Correlated versus Uncorrelated Stripe Pinning: the Roles of Nd and Zn Co-Doping}
\author{C. Morais Smith$^{a,b}$, N. Hasselmann$^b$, and A. H. Castro Neto$^c$}
\address{$^{a \,}$Institut de Physique Th\'eorique, Universit\'e de Fribourg, P\'erolles, CH-1700 Fribourg, Switzerland \\
$^{b \,}$I Institut f{\"u}r Theoretische Physik, Universit{\"a}t 
Hamburg, D-20355 Hamburg, Germany \\ 
$^{c \,}$Dept. of Physics, University of California, Riverside, CA, 92521, USA}

\widetext
\date{\today}
\maketitle
\tightenlines
\widetext
\advance\leftskip by 57pt
\advance\rightskip by 57pt

\begin{abstract}
We investigate the stripe pinning produced by Nd and Zn 
co-dopants in cuprates via a renormalization group approach.
The two dopants play fundamentally different roles in the 
pinning process. While Nd induces a correlated
pinning potential that traps the stripes in a flat phase and suppresses 
fluctuations, Zn pins the stripes 
in a disordered manner and promotes line meandering.
We obtain the zero temperature phase diagram and 
compare our results with neutron scattering data. A good agreement
is found between theory and experiment. 
\end{abstract}

]
\narrowtext

The existence of stripes in doped Mott insulators has engendered a great
debate recently. While the presence of stripes in the manganites and nickelates 
has been firmly established, uncertainties remain concerning whether they are
present in the cuprates. 
In manganites and nickelates stripes are static and can be 
easily observed.\cite{Mori,TranNi} In cuprates, on the other hand, 
they form a collective fluctuating state and their detection is more subtle.
Co-doping of cuprates has been extremely important for  
unveiling the modulated charge states. However, the inclusion of  
co-dopant usually reduces the critical superconducting temperature, 
$T_c$, raising doubts about the coexistence of superconductivity and 
the striped phase.\cite{TranCu,Koik1} 
The first experimental detection of stripes in the cuprates was achieved 
in a Nd co-doped compound La$_{2-x-y}$Nd$_y$Sr$_x$CuO$_4$. For $y=0.04$ 
and $x=0.12$ Tranquada {\it et al.}\cite{TranCu} found that the 
commensurate magnetic peak at ${\bf Q} = (\pi/a,\pi/a)$ splits by a 
quantity $\delta$, giving rise to four incommensurate peaks. In addition, the
Bragg peaks split by $2 \delta$, indicating that the charges form domain walls and that
the staggered magnetization undergoes a $\pi$-phase shift when crossing them.  
The study of co-doped cuprates has also involved other elements,  
such as Zn, Ni, Fe, Co, etc.\cite{Koik1,Xiao,Koik2,Koik3}

In this paper we study the problem of co-doping within the stripe scenario by
performing a renormalization group calculation on a model of quantum elastic 
strings under the influence of lattice and disorder potentials. We determine 
the zero temperature pinning phase diagram and
compare it with experimental data. The different roles played by rare earth 
(Nd, Eu) and planar (Zn, Ni) impurities led us to 
establish a parallel between the stripe- and vortex-pinning
problem in high-$T_c$ superconductors.

By doping the antiferromagnetic insulator La$_{2}$CuO$_4$ with Sr, i.e.,
by replacing La$^{3+}$ with Sr$^{2+}$, charge carriers are introduced 
into the CuO$_2$ planes. In the stripe scenario the carriers, instead
of forming a homogeneous quantum fluid, arrange themselves into a
highly anisotropic charge-modulated state with one-dimensional (1D)
characteristics. The ionized
Sr dopants are a source of disorder since 
they are located randomly in the neighborhood of the CuO$_2$ planes.
Hence, the number of holes is intrinsically
connected with the number of pinning centers, 
and the stripes can be collectively
pinned by these point-like impurities. 
Despite the correspondence between the number of impurity pinning
centers (Sr) and the charge carriers (holes) within the CuO$_2$ planes, it
is experimentally possible to control these two
parameters independently. By co-doping the superconducting material with Nd or
Zn, for instance, one can alter the disorder without modifying the 
number of charge carriers.\cite{TranCu,Koik1,Koik2,Koik3}  
On the other hand, by 
growing the superconducting film over a ferroelectric substrate and using an 
electrostatic field as a control parameter, the number of charge carriers in the 
plane can be increased for a fixed doping concentration.\cite{Ahn} 
Hence, the treatment of these two parameters independently is an important 
theoretical problem. 

Calculations of the pinning
energy within a model in which stripes are regarded as elastic
strings have shown that the problem can be described by the 
Collective Pinning Theory \cite{Gian}, with a critical Larkin pinning 
length $L_c \sim 100  \AA $ for doping of order $x \approx 10^{-2}$ 
(Ref.\ 10). 
Another possible source of pinning for the stripes is
lattice distortion such as the tilt of the oxygen octahedra.
We have recently studied the role played by lattice and dopants 
and have generated a phase diagram in terms of the incommensurability,
$\delta$, and the ratio between kinetic and elastic stripe energy, $\mu$ 
(this parameter measures the strength of quantum fluctuations).\cite{Nils}
Three different phases were identified: at large values of $\mu$ and $\delta$, 
the stripes form a collective fluctuating state or quantum membrane phase;
as $\mu$ is reduced the stripes become
pinned by the underlying lattice and decoupled from each other leading to
the so-called flat phase; finally, at small values of $\delta$ and $\mu$ disorder
becomes relevant and the system can be described in terms of a disordered phase.
In this paper we will generalize our earlier approach in order to incorporate
the differences between different co-dopants. 
 
In general, impurities will pin the stripes, leading to
the formation of a static charge order, which is usually accompanied by
a reduction of $T_c$. This statement holds for co-doping with several types
of impurities, such as Zn, Ni, Nd, Eu, independently of the intrinsic characteristics
of each dopant.\cite{TranCu,Koik1} Moreover, a special reduction of $T_c$ takes 
place when the effective number of charges in the CuO$_2$ plane is $n \sim 1/8$ 
(Refs.\ 5-7).
At this doping value, the striped structure becomes commensurate with
the underlying lattice and the effective pinning potential for collective motion
of the stripes is at a maximum.\cite{footnote}
A second important feature of doping within the stripe model is that the
average separation $L$ between neighboring stripes is not expected to change
upon co-doping if the substitution element has the same valence as the replaced
one. Therefore, co-doping will simply pin the stripes without changing their
overall number or separation. 
By replacing La$^{3+}$ with Nd$^{3+}$, for instance, one does {\it not} change 
the number of holes introduced into the plane.
The same argument holds if one replaces Cu$^{2+}$ by Zn$^{2+}$ or
Ni$^{2+}$. Hence, the average stripe separation $L$ and consequently the 
incommensurability $\delta = a/2L$ are not altered by the introduction of the
co-dopant, as is experimentally observed (and trivially inferred). 
\cite{TranCu,Yama,Kimu} We classify the pinning generated by co-dopants as  
{\it uncorrelated} or {\it correlated}. In the former
case the statistical mechanics of the stripes is characterized by line 
wandering,
whereas in the latter case the characteristic feature is localization.
The situation here is analogous to the case of a vortex line pinned by weak
point-like impurities (uncorrelated disorder) or by extended defects, like
1D screw dislocations or artificially produced columnar defects (correlated
disorder). For extended defects the pinning energy grows linearly with the
distance along the vortex for the case in which the vortex system is
properly aligned with the defect structure. This strong anisotropic pinning
is in contrast with the weak isotropic pinning produced by point-like
defects that compete with one another, leading to a square-root growth
of the pinning energy along the vortex line \cite{Gian} or the stripe.\cite{Cris}

We consider the transverse motion of stripes 
embedded in an antiferromagnetic background 
with lattice constant $a$. 
This is possible because the longitudinal and
transverse motions decouple due to magnetic confinement.\cite{sasha}
We restrict our studies to the underdoped regime, where the stripe-stripe
interaction is weaker than the interaction of each stripe with the lattice
and disorder pinning potentials. Hence, we assume that the 
stripe-stripe interaction is merely restricting the motion of one
stripe to a ``box'' of size $2L$ limited by the next neighboring stripes.  
This assumption simplifies the analysis to the case of
a single stripe which interacts with the lattice and impurity potentials. 
The phenomenological Hamiltonian describing the
system is
$$
H = \sum_n \left[ \frac{J}{2 a^2} 
\left( \hat{u}_{n+1} - \hat{u}_{n} \right)^2  
-2 t \cos\left(\frac{\hat{p}_n a}{\hbar} \right)
+  V_n(\hat{u}_n)\right],
$$
where $\hat{u}_n$ denotes the displacement of the $n$-th hole from its equilibrium
position, $|\hat{u}_n| < L$, 
$\hat{p}_n$ is the canonically conjugate momentum, $t$ is the hopping
parameter, $J$ is the stripe stiffness, and $V_n$ is a random pinning potential
with Gaussian average over the disorder ensemble,
$\left<V_n(u) V_{n'}(u') \right>_D = D \delta(u-u') \delta_{n,n'}$. The parameter
$D$ measures the strength of disorder. 

The calculations can be simplified by going to the continuum limit 
and introducing
replicas. The replicated zero temperature action reads \cite{Nils}
($\phi^i=\sqrt{\pi} u^i /a$)
\begin{eqnarray}
\label{action}
S^r & = & \sum_i S_0[\phi^i] 
+ \frac{G}{a} \sum_i \int_0^\infty d\tau \int dy 
  \ \cos \left( 2 \sqrt{\pi} \phi^i \right)
\nonumber \\ & &
+ \frac{D}{2 a L\hbar} \sum_{i,i'} \int dy \int_0^\infty d\tau
d\tau'
\nonumber \\ & &
  \ \cos \left[ 2 \sqrt{\pi} \delta  
\left(\phi^i(y,\tau)-\phi^{i'}(y, \tau')\right) \right],
\end{eqnarray}
with the Gaussian action $S_0$ given by
\begin{eqnarray}
S_0[\phi(y,\tau)]= \frac{\hbar}{2 \pi \mu}
\int_0^\infty d\tau \int dy
\left[ \frac{1}{c}\left(\partial_\tau \phi \right)^2
+ c \left(\partial_y \phi \right)^2 \right].
\nonumber
\end{eqnarray} 
The stripes are oriented along the $y$-direction and $\tau$ is  
imaginary time.
The free stripe velocity is $c=a \sqrt{2 t J}/ \hbar$ and the dimensionless
parameter $\mu=\sqrt{2t/J}$  measures the competition between
kinetic and confining energies.
The parameter $G$  accounts for the lattice 
effects and $i$ counts
the replicas. 
The one-loop RG equations for 
$G$, $D$, and $\mu$  
were obtained in Ref.\ 11. The phase diagram can
be divided into a flat phase (correlated pinning), a
disordered phase (uncorrelated pinning)
and a membrane phase. Here we focus on the pinned phases only. 
In these two phases the
single stripe approach is expected to be a reasonably good approximation
and we can obtain quantitative results on 
the effects of the different co-dopants.

{\it Correlated pinning:}
Let us first analyze the limit of vanishing disorder
$D=0$. The RG equations then read:
\begin{eqnarray}
\displaystyle{\frac{d}{d \ell}} \Gamma^2  &=&  2(2 - \pi \mu) \Gamma^2 \ ,  
\label{newrg1} \\
\displaystyle{\frac{d}{d \ell}} \mu^{-1}  &=& \displaystyle{\frac{1}{2}} \Gamma^2, 
\label{newrg2}
\end{eqnarray}
where $\Gamma =  \pi^{3/2} G a/(\hbar c)$.
Near the critical region,  we can
define the small
parameter $\epsilon_g = 2 - \pi \mu$ which measures the distance from the
critical line $\mu_{c1} = 2/\pi$, where the roughening transition 
(Kosterlitz-Thouless) takes place in the absence of co-doping.
Hence, $\mu^{-1} \approx (\pi/2) + (\pi/4) \epsilon_g$
and $d \epsilon_g / d \ell = (4/ \pi) d \mu^{-1} / d \ell$. Using (\ref{newrg2})
we obtain $d \epsilon_g / d \ell = (2/ \pi)\Gamma^2$, which can then be combined
with (\ref{newrg1}) yielding 
\begin{equation}
\displaystyle{\frac{d}{d \ell}} \left( \epsilon_g^2 - \frac{2}{\pi}\Gamma^2\right)  
= 0.
\label{finalrg}
\end{equation}
Eq. (\ref{finalrg}) implies that there is a transition at the critical value
$\epsilon_g^c = \pm \sqrt{2/\pi}\Gamma$. 
The critical value of $\mu_c(\Gamma)$ is thus
\begin{equation}
\mu_{c1}(\Gamma) = \frac{2}{\pi} + \sqrt{\frac{2}{\pi^3}} \Gamma.
\label{muc}
\end{equation}
For $\mu < \mu_{c1}(\Gamma)$ the stripes are pinned by the underlying lattice in
the so called ``flat phase''. The excitation spectrum is gaped and quantum
fluctuations are strongly suppressed. On the other hand, for 
$\mu > \mu_{c1}(\Gamma)$ the stripes are fluctuating freely. Consider a
fixed doping concentration $\delta$ for which the stripe system is in the
free phase for $\Gamma = 0$. Upon increasing the lattice parameter
$\Gamma$, the system moves along the thick line in Fig.\ 1 and eventually
enters the pinned ``flat phase'' after crossing the surface 
$\mu_{c1}(\Gamma)$.
\begin{figure}[t]
\unitlength1cm
\begin{picture}(5,5)
\epsfxsize=7cm
\centerline{\epsffile{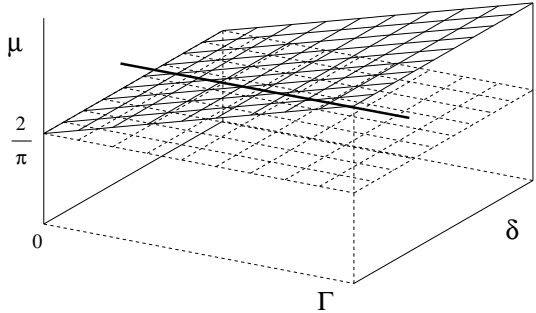}}       
\end{picture}
\vspace{0.1cm}
\caption[]{\label{topopicb} Pinning phase-diagram of the striped phase in the 
presence of correlated pinning.}
\label{phasel}
\end{figure}

This result can describe the effects
of Nd co-doping of the lanthanum cuprate. 
The introduction of Nd (or any other rare earth element) induces a structural 
transition
in the material from a low temperature orthorhombic (LTO) to a low temperature
tetragonal (LTT) phase, corresponding to a buckling of the oxygen octahedra. 
Hence, although the Nd randomly replaces the La atoms which are located
out of the plane, they indirectly produce a {\it correlated} pinning potential
along the copper lattice that will act to pin the stripes in the so called flat
phase, strongly suppressing thermal or quantum fluctuations. 
This is analogous
to the case of pinning of vortices by artificially introduced columnar defects.
Notice, however, that in the vortex-problem the columnar defects are randomly
distributed, whereas here the ``correlated'' pinning potential is actually
a periodic lattice potential, which is enhanced by the tilting of the oxygen
octahedra within the LTT phase. Despite of this difference,
the analogy is helpful, because it emphasizes the linear character
of the pinning potential in both problems.

Since the Nd pins the stripe in an ordered configuration, we expect the 
width of the incommensurate (IC) peaks measured by neutron scattering to be 
{\it reduced} and the 1D behavior to be reinforced by co-doping. These 
conclusions are supported by experimental data: ARPES \cite{Shen} and Hall
transport \cite{Noda} measurements of La$_{2-x-y}$Nd$_y$Sr$_x$CuO$_4$ 
show strong evidence for a 1D striped structure in the underdoped
regime. Besides, neutron scattering data taken for both compounds 
(with and without Nd) indicate a reduction of the half width
at half maximum (HWHM) IC peak in the presence of Nd for all the investigated
Sr compositions.\cite{TranCu,Yama}   
This is an indication that lattice pinning and hence commensuration
effects are enhanced through Nd doping.

\begin{figure}[t]
\unitlength1cm
\begin{picture}(5,5)
\epsfxsize=7cm
\centerline{\epsffile{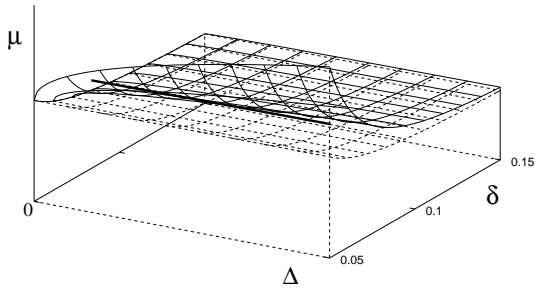}}       
\end{picture}
\vspace{0.1cm}
\caption[]{\label{topopica} Pinning phase-diagram of the striped phase in the 
presence of uncorrelated pinning. At low doping values the stripes are pinned
in a fuzzy phase, whereas above the critical surface they are fluctuating freely.}
\label{phased}
\end{figure}

{\it Uncorrelated pinning:} 
A completely different scenario is presented for the 
Zn doping case: Zn$^{2+}$ replaces Cu$^{2+}$ directly on the CuO$_2$ planes. They
are located {\it randomly} and act as uncorrelated point-like
pinning centers in a way very similar
to the pinning of vortices by oxygen vacancies. Again, they do not alter the
position of the IC peaks observed in neutron scattering, since
they do not change the hole density. 
Moreover, the randomly distributed Zn atoms induce stripe meandering and
pin the stripe in a fuzzy phase, similar to Sr doping.
The RG equations in the limit of negligible lattice pinning but relevant
disorder are \cite{Nils}
\begin{eqnarray} 
\displaystyle{\frac{d}{d \ell}} \Delta  
&=&  (3- \gamma \mu) \Delta \ , \\
\displaystyle{\frac{d}{d \ell}} \mu  &=&   
- \displaystyle{\frac{1}{2}}\mu^2 \Delta,
\label{disrg}
\end{eqnarray}
where $\gamma = 2 \pi \delta^2$ and $\Delta= 4 \pi^2 D \delta^2 a^2/
(\hbar^2 c^2 L)$. Close to the critical region we
define  $\epsilon_d = 3 - \gamma \mu$, and following a similar procedure
as done for the lattice pinning case, we obtain
\begin{equation} 
\displaystyle{\frac{d}{d \ell}} \left(\Delta - \frac{\gamma}{9} \epsilon_d^2 \right)  
=  0,
\end{equation}
indicating that $\Delta - (\gamma /9) \epsilon_d^2$ is preserved 
under the RG flow. The transition then happens at the critical value
\begin{equation}
\mu_{c2} = \frac{3}{2 \pi \delta^2} + \frac{3 \sqrt{\Delta}}{(2 \pi \delta^2)^{3/2}} \, .
\label{muc2}
\end{equation}
The corresponding phase diagram is shown in Fig.\ \ref{phased}.
The thick
line indicates how the system undergoes a transition for a constant $\delta$
from a ``free phase'' $(\mu > \mu_{c2})$ at $\Delta = 0$ to a ``fuzzy 
phase''  $(\mu < \mu_{c2})$ at finite $\Delta$.
An inspection of the phase diagram indicates that uncorrelated
pinning is more relevant at low doping values. Hence, we expect the
effects of Zn co-doping to decrease with doping.
Moreover, in contrast to Nd doping, Zn pinning destroys the 1D behavior and 
{\it increases}
the width of incommensurate neutron scattering peaks implying that 
the stripes are pinned within a broader region. 
This is indeed observed experimentally.\cite{Yama,Kimu,Hiro}
Neutron scattering measurements in La$_{2-x}$Sr$_x$Cu$_{1-y}$Zn$_y$O$_4$ 
for $y = 0.012$ and $x = 0.14$ show that Zn produces no relevant effect
and that the width
$\kappa_s$ of the IC peaks remains practically unaltered. 
The IC peak width for La$_{1.85}$Sr$_{0.15}$CuO$_4$ is
$\kappa_s = 0.020 \pm 0.006$ \AA ($E$ = 8 meV, $T$ = 8 K) (see Ref.\ 13) 
and the Zn-doped compound with a similar Sr-concentration ($x = 0.14$ and 
$y = 0.012$) displays the same features within the experimental error bars: 
$\kappa_s = 0.014 \pm 0.002 \AA$ for $E$ = 5 meV, $T$ = 10 K (Ref.\ 18).
The scenario changes quite a bit in the underdoped regime. For $x = 0.12$
and $y = 0.03$ (a composition for which superconductivity is completely
suppressed) the elastic IC peaks were observed at the same position
as for the Zn-free material, but $\kappa_s$ was increased
due to the doping: $\kappa_s < 0.005$ \AA$^{-1}$ for the Zn-free material, 
whereas $\kappa_s = 0.013(1)$ \AA$^{-1}$ for $y = 0.03$
(Ref.\ 14), reflecting the random character of the pinning centers. Although 
the commensurability at $x = 0.12$ makes this point special, we
expect this trend (increase of $\kappa_s$ upon Zn-doping) to continue,
especially at lower values of $x$. 
We emphasize that the pinning energy grows sub-linearly with the length of the 
stripe in the case of uncorrelated disorder.\cite{Cris} 
Hence, the Zn-pinned phase is analogous to
the vortex-glass phase discussed in the context of vortex creep.\cite{Gian}

In conclusion, we have shown that the main experimental features of co-doping
in cuprates can be understood within models of lattice-pinned or disorder-pinned
stripes. We divide the co-dopants into two classes: 
those which produce correlated and and those which produce 
uncorrelated pinning. 
Correlated pinning is produced through rare earth co-doping. The problem is
analogous to the pinning of vortices by columnar defects or screw dislocations.
In this case the stripes are
pinned in a flat phase and the fluctuations are strongly suppressed. The 
effective stripe width is reduced and consequently the IC neutron scattering
peaks become sharper after the introduction of the co-dopant. 
On the other hand, in-plane Zn- or Ni-doping  
provides randomly distributed point-like pinning centers, similar to the
oxygen vacancies in the vortex-creep problem. 
Within our model, in which the stripe is regarded as a quantum elastic
string, the effect of randomness is to ``disorder'' the
string, increasing the effective stripe width and broadening the IC peaks.
We expect this kind of pinning to be relevant only at low doping, 
as indicated in the phase diagram shown in Fig.\ \ref{phased},
in agreement with the experimental results. 

We are indebted with G.\ Blatter, D.\ Baeriswyl, A.\ O.\ Caldeira, R.\ Noack, 
S.\ Uchida, and K.\ Yamada for fruitful discussions. 
N.~H. is financially supported by the Graduierten Kolleg 
``Physik nanostrukturierter Festk\"orper.'' 
A.~H.~C.~N. acknowledges support from a LANL CULAR grant.


\begin{references}
\bibitem{footnote} Notice that the commensurability or incommensurability of the
stripe array with the underlying copper lattice is a different issue than the
incommensurability $\delta$ measured with neutron scattering. When $\delta \propto x
= 1/8$ the stripe array is commensurate with the lattice.
\end{references}
\end{document}